\documentclass[twocolumn,showpacs]{revtex4}
\usepackage{amsmath}
\usepackage[]{graphicx}

\begin{document}

\title{Auxetic properties of polycrystals}


\author{Cz. Jasiukiewicz}
\author{T. Paszkiewicz}
\email{tapasz@prz.edu.pl}
\author{S. Wolski}



\affiliation{Rzeszow University of Technology,\\ ul. W. Pola 2,  35-959 Rzeszow, Poland
  }


\begin{abstract}
Young's and shear moduli and Poisson's ratio of polycrystalline solids consisting of 2D quadratic and 3D cubic randomly oriented grains of the same size and shape is studied. Considered polycrystals are initially unstrained. It is shown that for such polycrystals the division of the mechanical stability regions into areas of various auxeticity properties is different than for monocrystalline solids. In particular the regions of complete auxeticity enlarge.  
\end{abstract}
\maketitle                  

\section{Introduction} 
\label{sc:introd}
In our recent papers we considered mechanical characteristics of all 2D symmetry systems \cite{japawol1} and of 3D crystalline structures of high and middle symmetry \cite{pawol1}. In particular we derived explicit expressions for Young's $E$ and shear $G$ moduli as well as for Poisson's ratio $\nu$ depending on directional cosines of angles between directions $\bf{n}$ of the load and the direction $\bf{m}$ of the lateral strain, respectively, and directions of the crystalline symmetry axes.

Our particular attention was paid to 2D \cite{japawol2} and 3D cubic \cite{pawol2} structures. All initially unstrained quadratic and cubic materials are characterized by three parameters $s_{1}$, $s_{2}$ and $s_{3}$ belonging to a half-infinite ($s_{1}>0$) prisms with (stability) triangles (ST) in the base lying in the ($s_{2}$,$s_{3}$)-plane. 

Among mechanical characteristics, Poisson's ratio $\nu$ is particularly interesting because the phenomenon of a negative $\nu$, i.e. solids expanding transversely to an applied tensile  stress, is counterintuitive. Elastic materials with negative $\nu$ are termed auxetics. Materials with negative $\nu$ for all pairs of vectors $\bf{n}$, $\bf{m}$ are $\it{complete}$ auxetics. Materials are auxetics if there exists pairs of $\bf{n}$, $\bf{m}$ for which $\nu<0$. Materials are non-auxetics if $\nu>0$ for all pairs of vectors $\bf{n}$, $\bf{m}$. In papers \cite{japawol2} and \cite{pawol2} we established regions of ST in which crystalline quadratic and cubic materials are completely auxetic, auxetic and non-auxetic. One may expect that complete auxetics are particulary promising from the standpoint of technology.

Many technically important materials are polycrystalline. In view of potential applications of crystalline auxetic materials, the question how polycrystallinity influences the auxetic properties deserves some attention. 

The elastic properties of polycrystalline materials depend on the single-crystal elastic constants of the crystallites which build up the polycrystal and on the manner in which crystallites are connected. In most cases exact orientations, shapes and connections are not known. These gaps in knowledge can be overcame  with the help of orientation and grain shape distribution functions. Generally, one has to resort to more or less realistic assumptions (cf. \cite{hirsekorn}). We assume that the grains have the same shape and volume and that their orientations are completely random.  
\section{Mechanical characteristics of monocrystalline 3D and 2D cubic elastic materials}
\label{sc:mech_char_mono}
\subsection{Relations between the stiffnesses and compliances and components of vectors $\bf{n}$ and $\bf{m}$}
\label{sc:relations}
Mechanical properties of crystalline materials depend on the components of the stiffness tensor $\mathbf{C}$ or on the components of compliance tensor $\mathbf{S}$. In the coordinate system related to the crystalline axes these tensors have the minimal number of independent components that depend on the particular crystalline symmetry. In the case of cubic and quadratic (i.e. 2D cubic) materials, these independent components are $C_{11}$, $C_{12}$ and $C_{66}$ and $S_{11}$, $S_{12}$ and $S_{66}$ respectively (cf. \cite{wallace}, \cite{musgrave}). The components $C_{IJ}$ and $S_{IJ}$ ($I,J=1,2,3,\ldots,6$) are related. For quadratic materials \cite{japawol1}
\begin{equation}
S_{11}=\frac{C_{11}}{C_{11}^{2}-C_{12}^{2}},\, S_{12}=-\frac{C_{12}}{C_{11}^{2}-C_{12}^{2}},\, S_{66}=\frac{1}{C_{66}}.
\end{equation} 
For cubic materials \cite{nye}
\begin{eqnarray}
\label{compliances_vs_stiffnesses}
	S_{11}=\frac{C_{11}+C_{12}}{\left(C_{11}+2C_{12}\right)\left(C_{11}-C_{12}\right)},\nonumber\\  	  		S_{12}=-\frac{C_{12}}{\left(C_{11}+2C_{12}\right)\left(C_{11}-C_{12}\right)},\, S_{44}=\frac{1}{C_{66}}.
\end{eqnarray}

The direction $\bf{n}$ of applied tension and direction  $\bf{m}$ in which the lateral expansion/contraction is measured are mutually perpendicular, i.e. ${\bf{nm}}=0$. For quadratic materials 
\begin{eqnarray}
\label{eq:m_vector1}
n_{1}=m_{2}=\cos\varphi,\nonumber\;\\  
n_{2}=-m_{1}=\sin{\varphi}.
\end{eqnarray}
In the case of cubic materials \cite{aouni_wheeler}
\begin{eqnarray}
\label{eq:n_vector}
n_{1}=\cos{\alpha}\cos{\varphi}\cos{\theta}-\sin{\alpha}\sin{\theta},\nonumber \;\\  
n_{2}=\cos{\alpha}\cos{\varphi}\sin{\theta}+\sin{\alpha}\cos{\theta},\nonumber \;\\ 
n_{3}=-\cos{\alpha}\sin{\varphi},
\end{eqnarray}
and 
\begin{eqnarray}
\label{eq:m_vector}
m_{1}=\sin{\varphi}\cos{\theta},\nonumber\;\\  
m_{2}=\sin{\varphi}\sin{\theta},\nonumber\;\\ 
m_{3}=\cos{\varphi},
\end{eqnarray}
where
\begin{equation}
 -\pi<\alpha\leq\pi,\, 0<\varphi\leq\pi,\, 0<\theta\leq\ 2\pi.
\label{eq:angle_inequalities}
\end{equation}

\subsection{Mechanical characteristics of 2D and 3D cubic materials}
\label{sc:mech_char}
In the case of quadratic materials we obtained \cite{japawol1}
\begin{eqnarray}
E^{-1}_{q}(\textbf{n})=s_{J}/2 + s_{L}\left(n^{2}_{1}-n^{2}_{2}\right)/2 \nonumber\\ 
+ 2s_{M}n^{2}_{1}n^{2}_{2}\geq 0,\label{E-quadr}
\end{eqnarray}
\begin{eqnarray}
\left[4G_{q}(\textbf{m},\textbf{n})\right]^{-1}=s_{L}\left(m_{1}n_{1}-m_{2}n_{2}\right)^{2}/2 \nonumber \;&&\\ 
+s_{M}\left(m_{1}n_{2}+m_{2}n_{1}\right)^{2}/2\geq 0,\label{g-quadr}&&
\end{eqnarray}
\begin{eqnarray}
-\frac{\nu_{q}(\textbf{m},\textbf{n})}{E_{q}(\textbf{n})}=s_{J}/2 + s_{L}\left(m_{1}^{2}-m_{2}^{2}\right)\left(n_{1}^{2}-n_{2}^{2}\right)/2\nonumber\\ 
+2s_{M}m_{1}m_{2}n_{1}n_{2}.\nonumber\\ \label{nu-quadr}
\end{eqnarray}

In the case of cubic materials in paper \cite{pawol1} we obtained the familiar formulas \cite{nye} 
\begin{eqnarray}
E^{-1}_{c}(\textbf{n})=\left[\left(s_{J}-s_{M}\right)/3+s_{L}\right]\nonumber \;\\ 
+\left(s_{M}-s_{L}\right)p(\textbf{n}),\label{E-cub}\\ 
-\frac{\nu_{c}(\textbf{m},\textbf{n})}{E_{c}(\textbf{n})}=\left(s_{J}-s_{M}\right)/3\nonumber \;\\ 
+\left(s_{M}-s_{L}\right)P(\textbf{m},\textbf{n}),
\label{nu-cub}\\ 		
\left[4G_{c}(\textbf{m},\textbf{n})\right]^{-1}=s_{L}/2+\left(s_{M}-s_{L}\right)P(\textbf{m},\textbf{n}),
\label{g-cub}
\end{eqnarray}
where 
\begin{eqnarray}
p\left(\textbf{n}\right)=\sum_{i=1}^{3}n_{i}^{4},\label{p-function}\\
P\left(\textbf{m},\textbf{n}\right)=P\left(\textbf{n},\textbf{m}\right) =\sum_{i=1}^{3}\left(m_{i}n_{i}\right)^{2}.
\label{eq:P-function}
\end{eqnarray}

The coefficients $s_{J}$, $s_{L}$ and $s_{M}$ are the eigenvalues of the compliance tensor $\mathbf{S}$. They are related to the eigenvalues of the stiffness tensor $\mathbf{C}$, namely $s_{u}=c_{u}^{-1}$ ($u=J,L,M$). For quadratic materials \cite{japawol1}
\begin{equation}
s_{J}=S_{11}+S_{12},\, s_{L}=\left(S_{11}-S_{12}\right),\, s_{M}=S_{66}/2. 
\label{eq:eigenval_q}
\end{equation}
For cubic materials \cite{pawol1}
\begin{equation}
s_{J}=S_{11}+2S_{12},\, s_{L}=S_{66}/2,\, s_{M}=S_{11}-S_{12}. 
\label{eq:eigenv_c}
\end{equation}
\section{The stability conditions for cubic and quadratic materials}
\label{stability_cond}
Both stiffness and compliance tensors are positive, therefore their eigenvalues $s_{U}$, $c_{U}$ ($U=J,L,M$) are also positive. This yields the familiar inequalities (cf. \cite{wallace}, \cite{japawol1}), namely 
\begin{equation}
	A_{11},\; A_{66}>0, \, \left(A_{11}-A_{12}\right)>0\; (A=C,S),
\label{eq:stability_both}
\end{equation}
and
\begin{equation}
	A_{11}+(d-1)A_{12}>0,
	\label{eq:rem_cond}
\end{equation}
where $d$ depends on dimensionality $D$: $d=2$ for 2D and $d=3$ for 3D. We shall underline that the above inequalities hold only for initially unstrained crystals \cite{wallace}. 

For quadratic materials more reasonable choice of independent parameters was proposed in our paper \cite{japawol2}, and for cubic materials by Every \cite{every}
\begin{eqnarray}
	s_{1}=C_{11}+(d-1)C_{66}>0, \, s_{2}=\frac{C_{11}-C_{66}}{s_{1}},\nonumber \\ 
	 s_{3}=\frac{C_{11}-C_{12}-2C_{66}}{s_{1}}.
	\label{eq:every}
\end{eqnarray}
In terms of Every's parameters the inequalities (\ref{eq:stability_both}) and (\ref{eq:rem_cond}) read
\begin{equation}
	s_{1}>0, \, s_{2}<1,
	\label{eq:every_inequal}
\end{equation}
\begin{align}
	2D: \left(1-s_{2}+s_{3}\right)>0,\, \left(s_{2}-s_{3}/2\right)>0, \\
	3D: \left(1 + 2s_{2}\right)>\left|4s_{2}-3s_{3}-1\right|,\, \left(10s_{2}-6s_{3}\right)>1.
	\label{eq:every_inequal_remaining}
\end{align}
 
\begin{figure}[ht]
	\centering
		\includegraphics*[width=\linewidth]{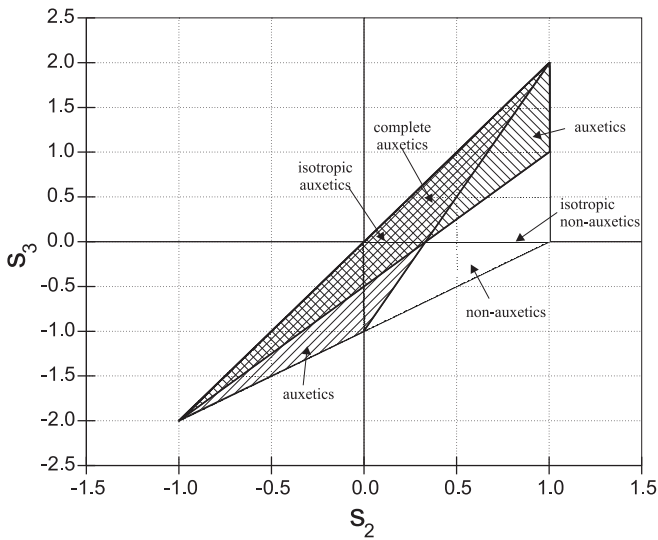}
	\caption{Division of the stability triangle for monocrystalline quadratic elastic materials into areas of various auxetic properties.}
	\label{fig:quadr_triangle}
\end{figure}

\begin{figure}[htb]
	\centering
		\includegraphics*[width=\linewidth]{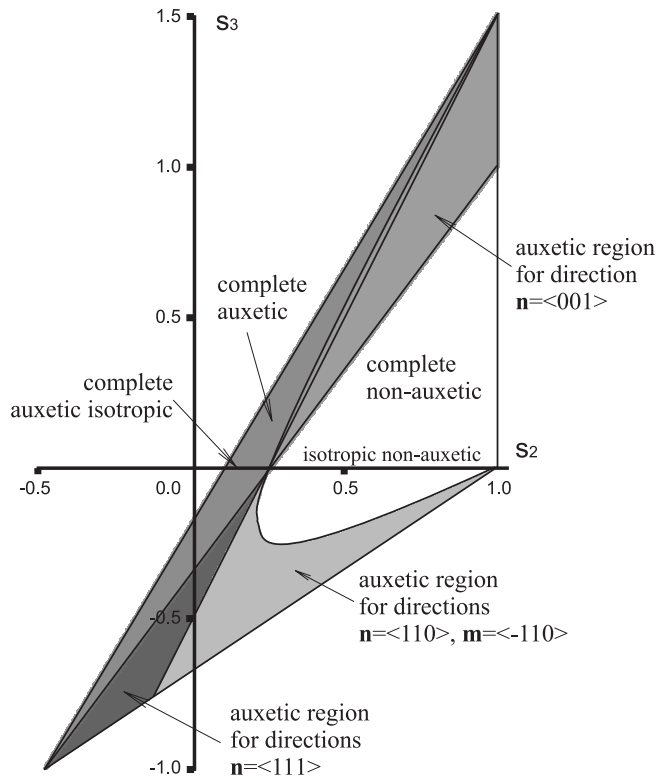}
	\caption{Division of the stability triangle for monocrystalline cubic elastic materials into areas of various auxetic properties.}
	\label{fig:cub_triangle}
\end{figure}

For both quadratic and cubic materials inequalities (\ref{eq:every_inequal})-(\ref{eq:every_inequal_remaining}) define stability regions in a form semi-infinite prisms ($s_{1}>0$) with two distinct triangles in the base. Both stability triangles (STs) are lying in the $(s_{2},s_{3})$-plane (Figs.   \ref{fig:quadr_triangle})and \ref{fig:cub_triangle}. Acoustic (\cite{every}, \cite{papru} and \cite{papruziel}) and elastic properties \cite{japawol2}, \cite{pawol2} of a particular initially unstrained cubic and quadratic material are characterized by $s_{1}$, which has dimensionality of stiffnesses, and two dimensionless parameters $s_{2}$ and $s_{3}$ belonging to the appropriate stability triangles. Isotropic materials are represented by points of the interval of the line $s_{3}=0$ belonging to the suitable ST. For acoustic and elastic characteristics the parameter $s_{1}$ is a scaling parameter. Regions of stability for symmetry systems other than cubic and isotropic as well as for the oblique symmetry system are defined in spaces of dimensions higher than 3. In the case of rectangular symmetry system the space of parameters can be effectively reduced to a 3D space \cite{japawol3}. 
\section{Procedure of averaging over direction of crystallites}
\label{averaging}
Following the assumptions made at the end of Sect. \ref{sc:introd} to find the mechanical characteristics of polycrystalline materials one should calculate mean values of expressions defining them. Consider quadratic materials and a function $F$ of angle $\varphi$, then in agreement with definitions (\ref{eq:m_vector1})
\begin{equation}
	\left\langle F\right\rangle=\frac{1}{2\pi}\int_{0}^{2\pi}d\varphi F(\varphi).
\label{eq:quadr_average}
\end{equation}

In the case of cubic materials one should consider function $F$ of angles $\alpha$, $\varphi$ and $\theta$ introduced in Sect. \ref{sc:relations}. The mean value $\left\langle F\right\rangle$ of $F(\alpha,\varphi,\theta)$ is defined as
\begin{equation}
\left\langle F\right\rangle=\frac{1}{8\pi^{2}}\int_{-\pi}^{+\pi}d\alpha\int_{0}^{\pi}{d\varphi}\sin\varphi\int_{0}^{2\pi}d\theta F(\alpha, \varphi, \theta).
\label{eq:cub_average}
\end{equation}
Since we consider the model of randomly oriented crystallites, the averaged mechanical characteristics are isotropic. This means that if $\left\langle \nu\right\rangle>0$, one deals with an isotropic non-auxetics, whereas if $\left\langle \nu\right\rangle<0$, a polycrystal is a complete isotropic auxetic. We conclude that in the case of considered models of polycrystals the division of the stability triangles simplifies. 

After averaging $\nu_{q}(\varphi)=\left(\nu_{q}(\varphi)/E_{q}(\varphi)\right)E_{q}(\varphi)$ we obtain 
\begin{equation}
	\left\langle \nu_{q}\right\rangle(s_{2},s_{3})=1-2\sqrt{\frac{\left(s_{2}-1\right)\left(s_{2}-s_{3}-1\right)}{\left(s_{2}+1\right)\left(s_{2}-s_{3}+1\right)}}.
	\label{nu_q_av}
\end{equation}
\section{Division lines for crystalline and polycrystalline quadratic and cubic materials}
\label{sc:div_lines}
Figs. 1 and 2 present the lines dividing the stability triangles into areas of various auxetic properties for monocrystalline quadratic and cubic materials. Since each part of STs has a finite area, the property of auxeticity is quite widespread.

Consider quadratic materials. The condition $\left\langle \nu_{q}\right\rangle=0$ yields an equation of dividing line
\begin{equation}
s^{(\nu)}_{3}=\frac{3s_{2}^{2}-10s_{2}+3}{3s_{2}-5}.	
\label{eq:quadr_div_line}
\end{equation}

Since $E_{q}({\bf{n}})\neq 0$ one can consider a different line defined by the condition $\left\langle \nu_{q}/E_{q}\right\rangle=0$ 
\begin{equation}
s^{(\nu_{q}/E_{q})}_{3}\left(s_{2}\right)=3s_{2}-2+\sqrt{3s_{2}^{2}-4s_{2}+2}.
\label{eq:quadr_nu-E_line}	
\end{equation}

\begin{figure}[htpb]
	\centering
		\includegraphics*[width=\linewidth]{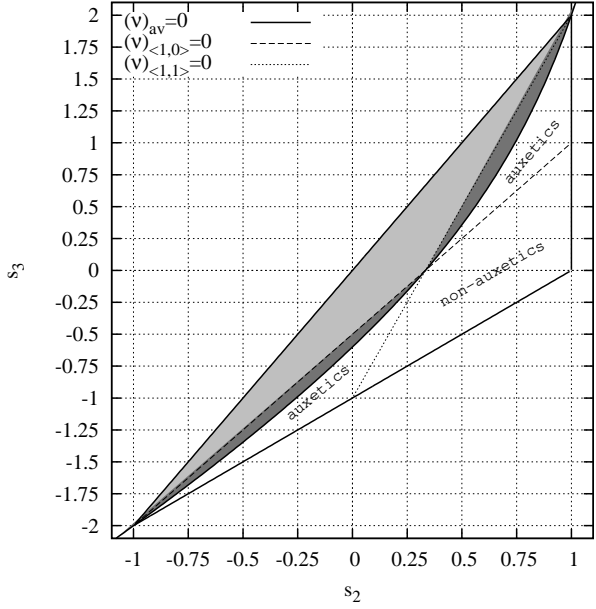}
	\caption{Comparison of division lines of ST for monocrystalline and polycrystalline quadratic elastic materials. Points of area with lighter shade  represent complete monocrystaline auxetics, whereas in the area with darker shade lie complete polycrystaline auxetics. }
	\label{fig:poly_quadr_triangle}
\end{figure}

Lines (\ref{eq:quadr_div_line}) and (\ref{eq:quadr_nu-E_line}) intersect at points $s_{2}=\pm 1$ and on the line $s_{3}=0$. Inspecting Fig.  \ref{fig:poly_quadr_triangle} we note that for both considered polycrystalline materials the area of complete polycrystalline auxetics is larger than for monocrystalline auxetics. This means that there exist cubic and quadratic auxetic materials that become complete auxetics in the polycrystalline phase.  

\begin{figure}[htpb]
	\centering
	\includegraphics*[width=\linewidth]{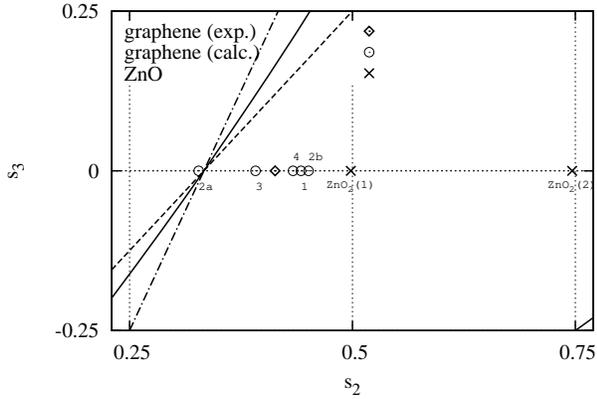} 
	\caption{Location of graphene on the isotropy line. The location of graphene-like {\rm ZnO} is also indicated. Graphene references: 1-3 -- \cite{michel_pr}, 4 and graphen exp. -- \cite{falkovsky}.  Reference for ZnO: \cite{tu}. Values of $s_{1}$ (in 10 GPa) for graphene: $s_{1}^{(1)}=56.2$, $s_{1}^{(2a)}=58$, $s_{1}^{(2b)}=71$, $s_{1}^{(3)}=53.7$, $s_{1}^{(4)}=120$, $s_{1}^{\rm (exp.)}=150$, 
for {\rm ZnO}: $s_{1}^{(1)}=97.5$, $s_{1}^{(2)}=65.1$. Results of Klintenberg et al \cite{klintenberg} for graphene are located between points 3 and 1.}
	\label{fig:graphene_on_triangle}
\end{figure}

Graphene has hexagonal symmetry, i.e. it is an isotropic material. Its elastic properties depend on two stiffnesses, namely $C_{11}$ and $C_{12}$. They were calculated by Falkovsky \cite{falkovsky}, Michel and Verberck \cite{michel_pr}, \cite{michel_pss} and Klitenberg et al \cite{klintenberg} and measured (cf. \cite{falkovsky}). Points on the $s_{3}=0$ isotropy line standing for these data are depicted in Fig. \ref{fig:graphene_on_triangle}. It is seen that in both monocrystalline and polycrystalline \cite{hibino} phases graphene is a non-auxetic elastic material. A point representing single {\rm ZnO} monolayer with graphene-like structure \cite{tu} is also set forth. Two points representing single {\rm ZnO} monolayer with graphene-like structure \cite{tu} is also set forth.

\begin{figure}[htpb]	\centering
	\includegraphics*[width=\linewidth]{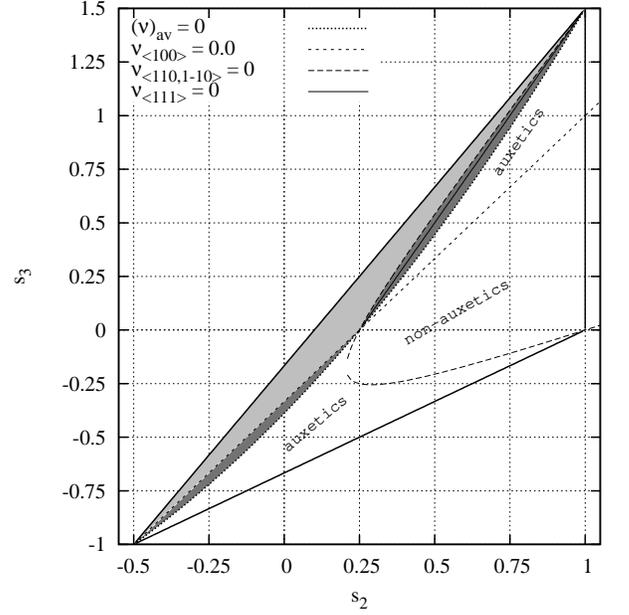}	
	\caption{Comparison of division lines of the ST for monocrystalline and polycrystalline cubic elastic materials.  Points of area with lighter shade  represent complete monocrystaline auxetics, whereas in the area with darker shade lie complete polycrystaline auxetics.}
	\label{fig:cubic_div_line}
\end{figure}

In the case of cubic materials numerical calculation gives the line of vanishing $\left\langle \nu_{c}\right\rangle=0$ which is presented in Fig. \ref{fig:cubic_div_line}.
\begin{figure*}[htb]%
\includegraphics*[width=\textwidth]{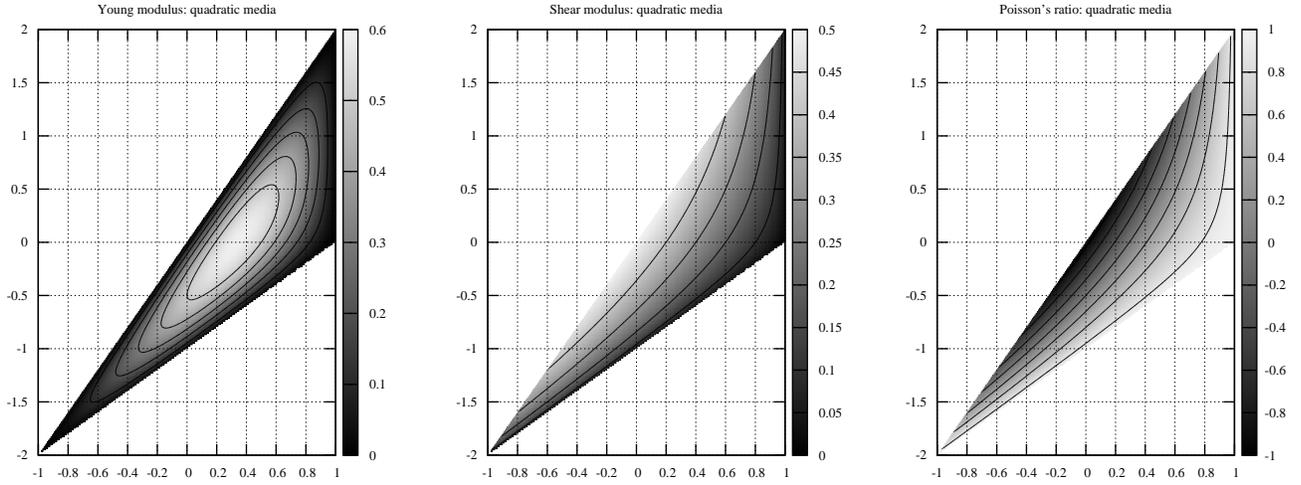}
\caption{Maps of $\left\langle E_{q}\right\rangle$, $\left\langle G_{q}\right\rangle$ and $\left\langle \nu_{q}\right\rangle$ for quadratic materials.}
\label{fig:quadratic-maps}
\end{figure*}
\begin{figure*}[htb]%
\includegraphics*[width=\textwidth]{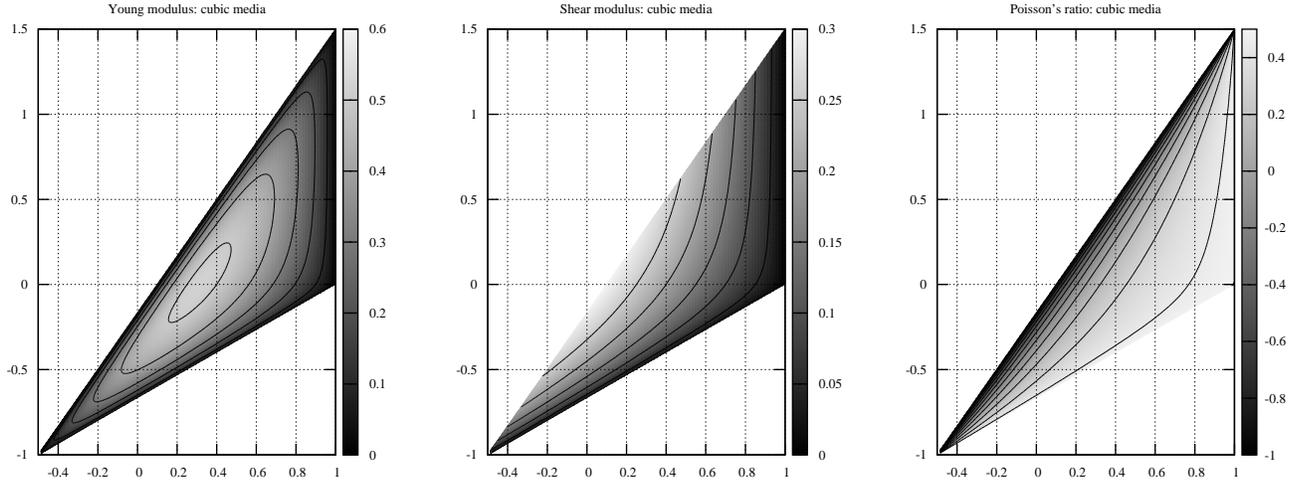}
\caption{Maps of $\left\langle E_{c}\right\rangle$, $\left\langle G_{c}\right\rangle$ and $\left\langle \nu_{c}\right\rangle$ for cubic materials.}
\label{fig:cubic-maps}
\end{figure*} 
\section{Mean values of mechanical characteristics of polycrystalline media}
\label{sc:mean-val-mechan}
In our previous papers we presented maps of $E$, $G$ and $\nu$ for quadratic materials \cite{japawol2}and cubic \cite{pawol2}. Since for monocrystalline media the mentioned characteristics are anisotropic, these maps refer to $\bf{n}$ and $\bf{m}$ directed along high symmetry crystalline directions. In the case of polycrystalline materials mechanical characteristics are isotropic. Their maps are presented in Figs. \ref{fig:quadratic-maps} and \ref{fig:cubic-maps}.

Calculating $\left\langle E_{q}\right\rangle$, $\left\langle G_{q}\right\rangle$ and $\left\langle \nu_{q}\right\rangle$ we obtained 
\begin{equation}
\left\langle E_{q}\right\rangle=\frac{4c_{J}c_{L}c_{M}}{\sqrt{\left[2c_{L}c_{M}+c_{J}\left(c_{L}+c_{M}\right)\right]^{2}-c_{J}^{2}\left(c_{L}-c_{M}\right)^{2}}},
	\label{eq:E-q-av}
\end{equation}
\begin{equation}
	\left\langle G_{q}\right\rangle=\sqrt{c_{L}c_{M}}/2,
	\label{eq:G-q-av}
\end{equation}
\begin{equation}
	\left\langle \nu_{q}\right\rangle=1-\frac{4c_{L}c_{M}}{\sqrt{\left[2c_{L}c_{M}+c_{J}\left(c_{L}+c_{M}\right)\right]^{2}-c_{J}^{2}\left(c_{L}-c_{M}\right)^{2}}}.
	\label{eq:nu-q-av}
\end{equation}

Table \ref{table} summarizes the properties of mean values of the mechanical characteristics and eigenvalues $c_{I}\; \left(I=J,L,M\right)$ on sides of ST. These properties align with calculated maps of $\left\langle E_{q}\right\rangle$, $\left\langle G_{q}\right\rangle$ and $\left\langle \nu_{q}\right\rangle$ (Figs. \ref{fig:quadratic-maps} and \ref{fig:cubic-maps}).
\begin{table}[h]
\caption{Values of $\left\langle E_{q}\right\rangle$, $\left\langle G_{q}\right\rangle$ and $\left\langle \nu_{q}\right\rangle$ on the sides of ST. J -- top side,  L -- vertical side, M -- bottom side.}
\begin{center}
\begin{tabular}
[c]{l|l|l|l|l}%
Side  & vanishing eigenvalue      & $\left\langle E_{q}\right\rangle$ & $\left\langle G_{q}\right\rangle$ & $\left\langle \nu_{q}\right\rangle$\\ \hline
J     &$c_{J}$             & 0                                 & $\sqrt{c_{M}c_{L}}/2$             & -1\\ \hline
L     &$c_{L}$             & 0                                 & 0                                 &  \;1\\ \hline
M     &$c_{M}$              & 0                                 & 0                                 &  \;1
\label{table}
\end{tabular}
\end{center}
\end{table}

Generally, for various points of ($s_{2},s_{3}$)-space, i.e. for various quadratic or cubic materials, the calculated mean values are different. Since on the line $s_{3}=0$ both  quadratic and cubic materials are isotropic, the maps for monocrystalline and polycrystalline materials should coincide along this line. This requirement is  fulfilled with a good accuracy. 
\section{Conclusions}
\label{sc:conclusions}
This paper has presented properties of the model of initially unstrained 3- and 2-dimensional cubic polycrystalline solids. We have assumed that grains have the same volume and shape, and are randomly oriented. We have shown that some monocrystalline auxetics become in the polycrystalline phase complete auxetics. 
	Generally, the averaged mechanical characteristics follow from the weighted integration over all orientations. Weighting function is the orientation distribution function which describes the texture of the polycrystal considered (eg. fibre textures). These more general models of polycrystalline solids certainly deserve further studies. 

\section*{Erratum}
\label{sc:erratum}

We noted three mistakes in article by Jasiukiewicz et al, physica status solidi vol. 245, issue 3, pp. 562-569: 
p. 564 formula for $r_{G}$ should read:
\begin{equation}
r_{G}=\sqrt{\left[ \frac{S_{66}+2S_{12}-S_{11}-S_{22}}{8}%
\right]^{2}+\left[ \frac{S_{16}-S_{26}}{4}\right]^{2}},  \nonumber
\end{equation}
The first of Eqs. (13) should read:
\begin{eqnarray}
	E_{o}^{-1}=S_{11}\cos^{4}\varphi + S_{22}\sin^{4}\varphi + 2S_{16}\cos^{3}\varphi\sin\varphi  \nonumber \\
	+ 2S_{26}\cos\varphi\sin^{3}\varphi + \left(2S_{12}+S_{66}\right)\cos^{2}\varphi\sin^{2}\varphi, \nonumber
\end{eqnarray}
p. 565 the definition of K should read: 
\begin{equation}
	K=\left(C_{11}-C_{12}-2C_{66}\right)/2. \nonumber
\end{equation}

\end{document}